\documentclass[%
 aip,
jcp,%
 amsmath,amssymb,
 reprint,%
 groupedaddress,
]{revtex4-1}
\usepackage{graphicx}
\usepackage{dcolumn}
\usepackage{bm}
\usepackage{xcolor}
\usepackage{siunitx}
\usepackage{epstopdf}

\newcommand{\Tr}{\mathrm{Tr}}
\newcommand{\tr}{\mathrm{tr}}
\newcommand{\rd}{\mathrm{d}}

\newcommand{\eu}[1]{\mathrm{e}^{#1}}

\usepackage{soul} 

\usepackage{braket}

\usepackage{dcolumn} 
\newcolumntype{d}[1]{D{.}{.}{#1}} 

\begin{document}

\title{A partially linearized spin-mapping approach for nonadiabatic dynamics.
\\ I.\ Derivation of the theory}

\author{Jonathan R. Mannouch}
\email{jonathan.mannouch@phys.chem.ethz.ch}
\author{Jeremy O. Richardson}%
\email{jeremy.richardson@phys.chem.ethz.ch}
\affiliation{Laboratory of Physical Chemistry, ETH Z\"{u}rich, 8093 Z\"{u}rich, Switzerland}

\date{\today}

\begin{abstract}
We present a new partially linearized mapping-based approach for approximating real-time quantum correlation functions in condensed-phase nonadiabatic systems, called spin-PLDM\@. Within a classical trajectory picture, partially linearized methods treat the electronic dynamics along forward and backward paths separately by explicitly evolving two sets of mapping variables. Unlike previously derived partially linearized methods based on the Meyer--Miller--Stock--Thoss mapping, spin-PLDM uses the Stratonovich--Weyl transform to describe the electronic dynamics for each path within the spin-mapping space; this automatically restricts the Cartesian mapping variables to lie on a hypersphere and means that the classical equations of motion can no longer propagate the mapping variables out of the physical subspace. The presence of a rigorously derived zero-point energy parameter also distinguishes spin-PLDM from other partially linearized approaches. These new features appear to give the method superior accuracy for computing dynamical observables of interest, when compared with other methods within the same class. The superior accuracy of spin-PLDM is demonstrated within this paper through application of the method to a wide range of spin-boson models, as well as to the Fenna--Matthews--Olsen complex.
\end{abstract}

\maketitle

\section{\label{sec:intro}Introduction}
The coupled dynamics of electrons and nuclei in molecular condensed-phase systems remains a challenging problem for computer simulation.\cite{Tully2012perspective,Stock2005nonadiabatic,Kapral2015QCL,Miller2001SCIVR} Nonadiabatic transitions can be induced when the electronic adiabatic states of a system become close in energy at some nuclear geometry, which results in a breakdown of the Born--Oppenheimer approximation. 

One approach is to use numerically exact wavefunction-based methods to compute the dynamical observables of interest. The advantage of such methods is that the obtained results are systematically improvable on increasing the size of the basis; they can also in principle describe nuclear quantum effects, such as tunnelling and zero-point energy. Of particular promise are methods, like the time-evolving block decimation (TEBD) technique,\cite{Vidal2003} which can even be applied to fairly large systems.\cite{Mannouch2018} However such methods often have restrictions on the type of systems that they can treat. For example, TEBD can only be efficiently applied to spatially linear systems with at most nearest-neighbour couplings. 

Due to the continuous nature of the nuclear subspace, the dynamics of such degrees of freedom are ideally suited to be performed using classical trajectories. The advantage of using such methods is that they are numerically cheap and can be easily applied to large, condensed-phase problems. Of these methods, the most popular are Tully's fewest-switches surface hopping,\cite{Tully1990hopping} which is heuristically motivated and hence cannot be rigorously derived\cite{Martens2020} and Ehrenfest dynamics,\cite{Ehrenfest1927,Mclachlan1964} which neglects all entanglement between the electronic and nuclear subsystems.\cite{Grunwald2009QCLE}

Recently, there has been a renaissance in mapping-based techniques,\cite{Cotton2013SQC,Huo2011densitymatrix,Huo2015PLDM,Miller2016Faraday,Liu2017,Kananenka2018,xin2019,Hsieh2012FBTS,Hsieh2013FBTS,Kelly2020,Liu2020,identity,FMO,linearized} which describe the dynamics of the electronic subsystem as well as the nuclei using classical trajectories within a continuous mapping space. Such methods have been shown to have superior accuracy compared to Ehrenfest dynamics, when calculating dynamical observables for a wide range of model systems. 
While these methods still treat the nuclear degrees of freedom classically and hence neglect nuclear quantum effects, these could be perhaps reintroduced by using ring-polymer based formulations.\cite{mapping,Ananth2013MVRPMD,Chowdhury2017CSRPMD,Tao2018isomorphic} Although none of these approaches will be able 
to describe quantum nuclear coherence effects,\cite{Miller2012} such effects are typically unimportant in condensed-phase systems.\cite{Miller2001SCIVR}

Historically, the vast majority of mapping-based approaches have used the so called Meyer--Miller--Stock--Thoss (MMST) mapping.\cite{Meyer1979spinmatrix,Stock1997mapping} In this scheme, the electronic subsystem is mapped onto a set of singly-excited harmonic oscillators. The nuclear and electronic dynamics are then described as an average over many classical trajectories. While trajectory-based methods using this mapping have been able to qualitatively reproduce the correct dynamics in a range of model systems, such methods may also exhibit significant problems. One of the most important of these is zero-point energy leakage, where zero-point energy can unphysically flow between the mapping harmonic oscillators as a result of the trajectories leaving the physical subspace during the classical dynamics.\cite{Mueller1998mapping,Kelly2012mapping} In other words, the trajectories can evolve into areas of phase space in the mapped system which do not correspond to a valid state in the electronic system. To alleviate this problem, related approaches have been suggested, such as reducing the zero-point energy parameter in the underlying theory,\cite{Mueller1999pyrazine} symmetric windowing\cite{Cotton2013SQC,Cotton2013mapping,Cotton2014ET,Cotton2015spin,Miller2015SQC,Miller2016Faraday,Cotton2016_2,Cotton2016SQC,Cotton2017mapping,Liang2018,Cotton2019,Cotton2019SQC} of trajectories, an MMST identity-trick\cite{identity,FMO,linearized} and using alternative classical mapping models.\cite{Liu2016,xin2019}

Recently a different form of mapping based on classical spin dynamics\cite{Meyer1979spinmatrix,Cotton2015spin} has been reformulated in terms of Stratonovich--Weyl kernels.\cite{spinmap,multispin} While this spin-mapping approach leads to exactly the same equations of motion for the trajectories as MMST mapping,\cite{Meyer1979spinmatrix} the Cartesian mapping variables are constrained to a hypersphere which is isomorphic with the phase space of the actual electronic subsystem. This means that unlike standard MMST mapping, spin-mapping does not need projection operators onto the physical subspace and trajectories no longer suffer from zero-point energy leakage. Also, when converted to Cartesian mapping variables, the spin-mapping Hamiltonian also has a different zero-point energy parameter to that of MMST mapping. Different zero-point energy parameters within MMST mapping have also been previously investigated by M{\"u}ller and Stock\cite{Mueller1999pyrazine} and have also previously been justified using analogies to spin by Cotton and Miller.\cite{Cotton2013mapping,Cotton2016SQC}

\begin{figure}
\includegraphics[scale=.3]{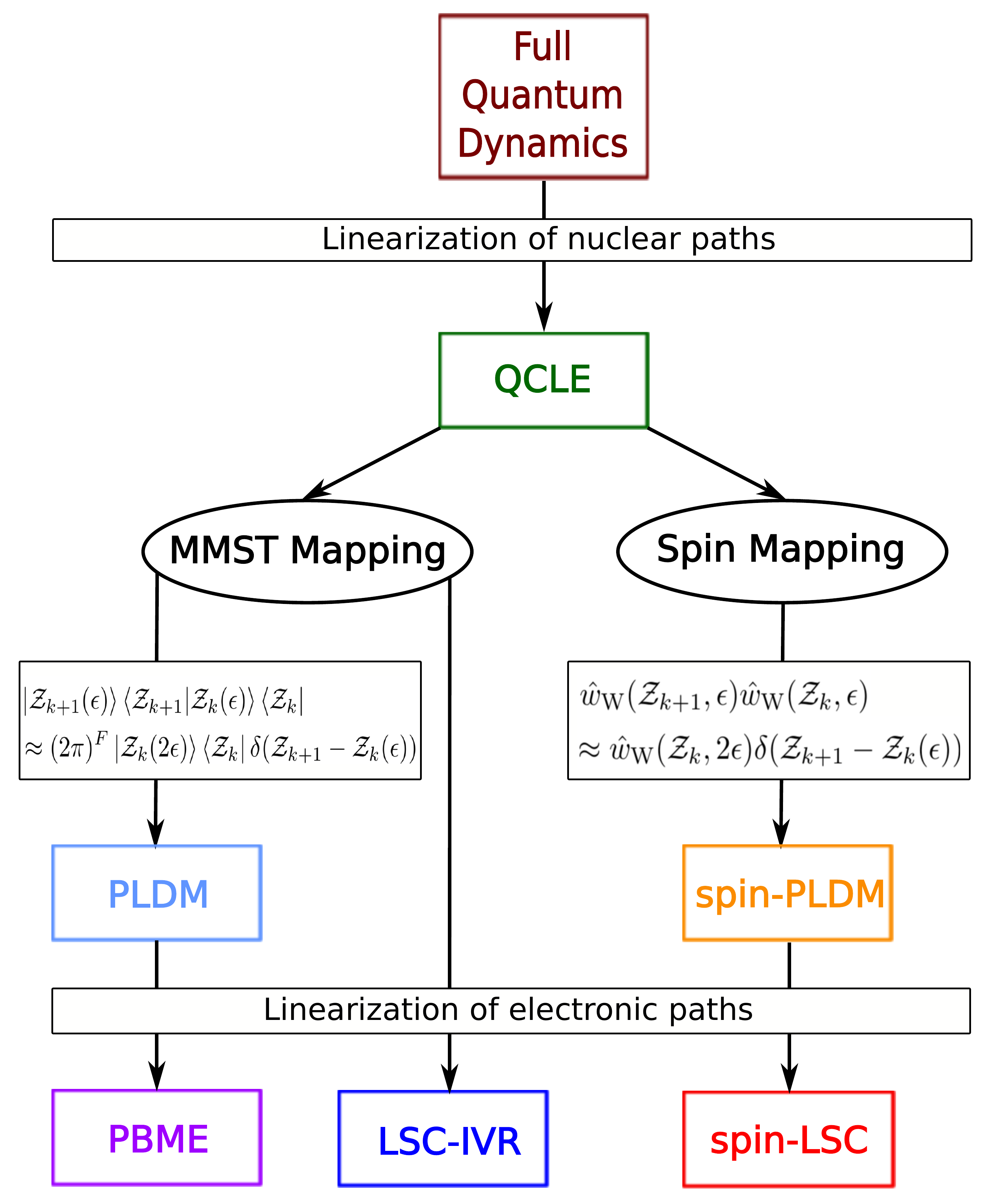}
\caption{Schematic illustrating the relationship between various mapping-based classical-trajectory techniques.}\label{fig:methods}
\end{figure}

When derived from a path-integral formalism, previous mapping-based techniques generally fall into one of two categories. Fully linearized methods\cite{Miller2001SCIVR,Sun1998mapping,Shi2004goldenrule,Kim2008Liouville,linearized} result from performing a linearization approximation to the difference between the forward and backward paths for both the electronic and nuclear degrees a freedom; a semiclassical approximation that is expected to be valid in the classical limit. In contrast, partially linearized methods\cite{Sun1997} result from only performing a linearization approximation for the nuclear paths and then treating the dynamics of the forward and backward electronic paths separately. Examples of partially linearized methods using MMST mapping are the partially linearized density matrix (PLDM) approach\cite{Huo2010,Huo2011densitymatrix,Huo2012MolPhys,Huo2015PLDM,Huo2012PLDM,*Huo2012_2,*Huo2013PLDM,Lee2016,Castellanos2017,Mandal2018quasidiabatic,Mandal2018_2,Mandal2019} and the forward-backward trajectory solution (FBTS).\cite{Hsieh2012FBTS,Hsieh2013FBTS,Kelly2020} While spin-mapping has already successfully improved the accuracy of so called fully linearized methods,\cite{spinmap,multispin} it has yet to be applied to partially linearized methods. Often partially linearized methods are better than their fully linearized counterparts, so we may optimistically expect a spin-mapping version of PLDM to be the best method of all. To aid the reader, a schematic illustrating the relationship of spin-PLDM to other mapping-based classical-trajectory techniques is given in Fig.~\ref{fig:methods}.

In this paper, after a brief review of the standard PLDM method based on MMST mapping, we use spin-mapping to derive a new partially linearized approach, which we call spin-PLDM\@. The method is derived generally for systems containing any number of electronic states by employing the successful framework introduced in Ref.~\onlinecite{multispin}. We show that such a method is able to more accurately reproduce dynamical observables associated with both the spin-boson model and Fenna-Matthews--Olsen (FMO) complex than other commonly used mapping-based approaches\@.  In Paper II,\cite{paper2} the spin-PLDM method is analyzed extensively to explain the reasons for its improved behaviour over other methods within the same class. 

\section{\label{sec:theory}Theory}
In this paper, 
we study nonadiabatic dynamics using the specific example of the familiar electronic-nuclear coupled system.
Naturally, the method we derive in this paper could also be applied to any other quantum-classical system. The Hamiltonian for such a system can always be written in the following form:
\begin{equation}
\label{eq:ham}
\hat{H} = \frac{\hat{p}^2}{2m} + V_{0}(\hat{x}) + \hat{V}(\hat{x}) ,
\end{equation}
where $V_{0}(\hat{x})$ is a state-independent potential function of nuclear configuration, $x$, whereas $\hat{V}(\hat{x})$ has an $F\times F$ matrix representation in the basis of electronic states, $\ket{1},\dots,\ket{F}$; these are uniquely partitioned so that $\tr[\hat{V}(\hat{x})]=0$. We define $F$ as the dimensionality of the electronic subsystem and $\tr[\cdot]$ as the partial trace over the electronic degrees of freedom. Both $x$ and $p$ are vectors of dimension $f$, which have been mass weighted such that all degrees of freedom have the same mass $m$. We set $\hbar=1$ throughout. 

Many dynamical quantities of interest can be written in the form of a real-time quantum correlation function:\cite{Nitzan}
\begin{equation}
\label{eq:corr}
C_{AB}(t)=\Tr\left[\hat{\rho}_{\text{b}}\hat{A}\,\eu{i\hat{H}t}\hat{B}\,\eu{-i\hat{H}t}\right] ,
\end{equation}
where $\hat{A}$ is an electronic operator effectively initializing the system at $t=0$, and $\hat{\rho}_{\text{b}}$ is the initial density matrix for the nuclear bath. These are known as factorized initial conditions. In addition, $\hat{B}$ is an electronic operator that measures the system at time $t$. Although we only consider the case where $\hat{A}$ and $\hat{B}$ are purely electronic operators in this paper, the theory could be easily extended for more general real-time correlation functions. Also, one can in principle go beyond factorized initial conditions using the more general density matrix $\hat{\rho}=\sum_{j}\hat{\rho}_\text{b}^{(j)}\hat{A}^{(j)}$.

Inserting complete sets of nuclear position eigenstates ($\ket{x_{0}}$, $\ket{x'_{0}}$ and $\ket{x_{N}}$) and electronic basis states ($\ket{\lambda}$ and $\ket{\mu}$) results in the following expression for the real-time correlation function: 
\begin{equation}
\begin{split}
&C_{AB}(t)=\sum_{\lambda,\lambda'}\sum_{\mu,\mu'}\braket{\lambda|\hat{A}|\lambda'}\braket{\mu'|\hat{B}|\mu}\int\rd x_{0}\rd x'_{0}\rd x_{N} \\ 
&\times\braket{x_{0}|\hat{\rho}_{\text{b}}|x'_{0}}\braket{\lambda',x'_{0}|\eu{i\hat{H}t}|\mu',x_{N}}\braket{\mu,x_{N}|\eu{-i\hat{H}t}|\lambda,x_{0}} ,
\end{split}
\end{equation}
where the unprimed coordinates are used for the forward propagation path, the primed coordinates for the backward propagation path and $x_{N}=x'_{N}$, because there is no nuclear operator applied at time $t$ within the correlation function $C_{AB}(t)$. One of the main challenges in evaluating this expression is obtaining an efficient expression for the forward and backward real-time propagators. When derived from a path-integral representation consisting of $N$ time steps, all mapping-based classical-trajectory approaches first implement a linearization approximation\cite{Miller2001SCIVR} for the nuclear degrees of freedom to avoid sampling complex phases. The linearization approach is also well-known for single-surface systems where it leads to classical Wigner dynamics.\cite{Hernandez1998,Wang1998,Miller2001SCIVR} Hence such a linearization approximation amounts to treating the dynamics of the nuclear degrees of freedom classically. 

To implement the linearization approach,\cite{Sun1997mapping,Sun1998mapping} the nuclear degrees of freedom must first be transformed to sum and difference coordinates, $\bar{x}_{k}=\tfrac{1}{2}(x_k+x_k')$ and $\Delta x_{k}=x_k-x_k'$ respectively, where $x_{k}$ ($x'_{k}$) is the nuclear coordinate at path-integral time-step $k$, associated with the forward (backward) propagation path. Then the approximation involves only retaining terms in the nuclear and electronic action up to first order in the difference coordinates, $\Delta x_{k}$. This results in the following approximate path-integral expression for the real-time correlation function:
\begin{equation}
\label{eq:corr_lin}
\begin{split}
&C_{AB}(t)\simeq\sum_{\lambda,\lambda'}\sum_{\mu,\mu'}\braket{\lambda|\hat{A}|\lambda'}\braket{\mu'|\hat{B}|\mu}\int\rd\bar{x}_{0}\rd\bar{p}_{0}\rho_{\text{b}}(\bar{x}_{0},\bar{p}_{0}) \\
&\times\!\!\int\!\!\frac{\rd\Delta p_{0}}{(2\pi)^{f}}\rd\bar{x}_{N}\!\!\prod_{k=1}^{N-1}\!\!\rd\bar{x}_{k}\frac{\rd\bar{p}_{k}}{(2\pi)^{f}}\rd \Delta x_{k}\frac{\rd \Delta p_{k}}{(2\pi)^{f}}T'_{[\lambda',\mu']}T_{[\mu,\lambda]}\eu{i\Delta S_{0}} ,
\end{split}
\end{equation}
where the initial nuclear density matrix is now described by its Wigner transform:
\begin{equation}
\label{eq:wigner}
\begin{split}
\rho_{\text{b}}(\bar{x}_{0},\bar{p}_{0})&= \\
\frac{1}{(2\pi)^{f}}&\int\rd \Delta x_{0}\Braket{\bar{x}_{0}+\frac{\Delta x_{0}}{2}|\hat{\rho}_{\text{b}}|\bar{x}_{0}-\frac{\Delta x_{0}}{2}}\eu{-i\bar{p}_{0}\Delta x_{0}} .
\end{split}
\end{equation}
Defined in this way, the Wigner transform of the initial nuclear density is normalized such that $\int\rd\bar{x}_{0}\rd\bar{p}_{0}\rho_{\text{b}}(\bar{x}_{0},\bar{p}_{0})=1$. Additionally the linear approximation to the difference of nuclear action along the two paths is given by:\cite{Hernandez1998,Sun1998mapping,Huo2011densitymatrix}
\begin{equation}
\label{eq:nuc_action_lin}
\begin{split}
\Delta S_{0}&\simeq\sum_{k=0}^{N-1}\left(\frac{\bar{x}_{k+1}-\bar{x}_{k}}{\epsilon}-\frac{\bar{p}_{k}}{m}\right)\epsilon\Delta p_{k} \\
&-\sum_{k=1}^{N-1}\left(\frac{\bar{p}_{k}-\bar{p}_{k-1}}{\epsilon}-F_{0}(\bar{x}_{k})\right)\epsilon\Delta x_{k} ,
\end{split}
\end{equation}
where $\epsilon=t/N$ is the path-integral time-step, 
\begin{equation}
\label{eq:force_indep}
F_{0}(\bar{x}_{k})=-\nabla V_{0}(\bar{x}_{k})
\end{equation}
is the state-independent nuclear force and $\nabla$ is the gradient, a vector of derivatives with respect to nuclear positions.

The electronic action is included within the definition of the electronic transition amplitude, $T_{[\mu,\lambda]}$. Retaining terms in the electronic action up to first order in the difference coordinates, $\Delta x_{k}$, the electronic transition amplitude for the forward and backward paths are approximated by:
\begin{subequations}\label{eq:trans_amp_lin}
\begin{align}
&T_{[\mu,\lambda]}\simeq\left\langle\mu\left|\eu{-i\hat{V}(\bar{x}_{N})\epsilon}\eu{-\tfrac{i}{2}\nabla\hat{V}(\bar{x}_{N-1})\epsilon \Delta x_{N-1}}\eu{-i\hat{V}(\bar{x}_{N-1})\epsilon}\cdots\right.\right. \notag\\ 
&\left.\left.\qquad\quad\qquad\qquad\qquad\qquad\times\eu{-\tfrac{i}{2}\nabla\hat{V}(\bar{x}_{1})\epsilon \Delta x_{1}}\eu{-i\hat{V}(\bar{x}_{1})\epsilon}\right|\lambda\right\rangle ,\label{eq:trans_amp_lina} \\
&T'_{[\lambda',\mu']}\simeq\left\langle\lambda'\left|\eu{+i\hat{V}(\bar{x}_{1})\epsilon}\eu{-\tfrac{i}{2}\nabla\hat{V}(\bar{x}_{1})\epsilon \Delta x_{1}}\cdots\right.\right. \notag\\
&\left.\left.\qquad\times\eu{+i\hat{V}(\bar{x}_{N-1})\epsilon}\eu{-\tfrac{i}{2}\nabla\hat{V}(\bar{x}_{N-1})\epsilon \Delta x_{N-1}}\eu{+i\hat{V}(\bar{x}_{N})\epsilon}\right|\mu'\right\rangle , \label{eq:trans_amp_linb}
\end{align}
\end{subequations}
These quantities contain two types of propagator: the electronic propagator, $\eu{\pm i\hat{V}(\bar{x}_{k})\epsilon}$, which describes the electronic dynamics; and the coupling propagator, $\eu{-i\nabla\hat{V}(\bar{x}_{k})\epsilon \Delta x_{k}/2}$, which due to its dependence on $\Delta x_k$ adds an electronic contribution to the nuclear force, as will be shown later.

With these approximate expressions for the nuclear action between the two paths [Eq.~(\ref{eq:nuc_action_lin})] and the transition amplitudes for the forward and backward paths [Eq.~(\ref{eq:trans_amp_lin})], the real-time correlation function would be an exact solution of the quantum-classical Liouville equation (QCLE).\cite{Kapral1999,Bonella2010,Shi2004QCLE} However, it is known that it is not possible to implement such an expression in an independent trajectory simulation.\cite{Kapral1999} The easiest way to further simplify the expression for the real-time correlation function is by introducing a mapping variable representation for the electronic transition amplitudes. In general this cannot be performed exactly within a classical-trajectory picture and hence different mapping-based classical-trajectory techniques differ by how they approximate the electronic transition amplitudes. We first consider the electronic transition amplitudes within the MMST mapping representation and then how these quantities are approximated within the standard partially linearized density matrix (PLDM) approach. We now set $\bar{x}\rightarrow x$ and $\bar{p}\rightarrow p$ throughout the rest of the paper. 
\subsection{\label{sec:pldm}The standard partially linearized density matrix approach}
The standard partially linearized density matrix approach (PLDM) \cite{Huo2011densitymatrix} represents the electronic transition amplitudes using the Meyer-Miller-Stock-Thoss mapping.\cite{Meyer1979spinmatrix,Stock1997mapping} Within this MMST mapping, the electronic basis states, $\ket{\lambda}$, are described by the single phonon excitation subspace of a set of $F$ harmonic oscillators, where $F$ is the size of the electronic system.\cite{Stock1997mapping} Within the standard PLDM/FBTS derivation, the electronic transition amplitudes are approximated within this MMST mapping space as follows. First, the coupling propagator is evaluated using a harmonic oscillator coherent state basis:\cite{Hsieh2012FBTS}
\begin{equation}
\label{eq:mmst_force_kernel}
\begin{split}
&\text{e}^{-\tfrac{i}{2}\nabla\hat{V}(x_{k})\epsilon \Delta x_{k}}= \\
&\frac{1}{2\pi}\int\rd\mathcal{Z}_{k}\,\ket{\mathcal{Z}_{k}}\eu{-\tfrac{i}{2}\nabla V_{\text{m}}(\mathcal{Z}_{k},x_{k})\epsilon \Delta x_{k}}\bra{\mathcal{Z}_{k}}+\mathcal{O}(\epsilon^{2}) . 
\end{split}
\end{equation}
where $\mathcal{Z}_{k}=\{Z^{(k)}_{1},Z^{(k)}_{2},\cdots,Z^{(k)}_{F}\}$ are the Cartesian mapping variables for the electronic subsystem and $V_{\text{m}}(\mathcal{Z}',x)$ is the (scalar) mapping representation of the traceless potential operator:
\begin{equation}
\label{eq:map_ham}
V_{\text{m}}(\mathcal{Z},x)=\tfrac{1}{2}\sum_{\lambda,\mu}\braket{\lambda|\hat{V}(x)|\mu}Z^{*}_{\lambda}Z_{\mu} .
\end{equation}
Eq.~(\ref{eq:mmst_force_kernel}) is exact when the $N\rightarrow\infty$ limit of the path-integral expression is taken. Second, the action of the electronic propagator on a harmonic oscillator coherent state is exactly described by evolving the Cartesian mapping variables under the following equations of motion:
\begin{subequations}
\label{eq:elec_eom}
\begin{align}
&\ket{\mathcal{Z}_{k}(\epsilon)}=\eu{-i\hat{V}(x_{k+1})\epsilon}\ket{\mathcal{Z}_{k}} , \\
&\frac{\rd Z^{(k)}_{\lambda}}{\rd\epsilon}=-i\sum_{\mu}\braket{\lambda|\hat{V}(x)|\mu}Z^{(k)}_{\mu} .
\end{align}
\end{subequations}

Together, Eqs.~(\ref{eq:mmst_force_kernel}) and (\ref{eq:elec_eom}) lead to an expression for the electronic transition amplitudes [Eq.~(\ref{eq:trans_amp_lin})] which contain an integral over Cartesian mapping variables at each time-step. In order to describe the electronic transition amplitudes approximately using continuous trajectories, both standard PLDM and FBTS implement the following approximation involving the overlap of harmonic oscillator coherent states:\cite{Hsieh2012FBTS}
\begin{equation}
\begin{split}
\label{eq:coherent_delta}
\ket{\mathcal{Z}_{k+1}(\epsilon)}&\braket{\mathcal{Z}_{k+1}|\mathcal{Z}_{k}(\epsilon)}\bra{\mathcal{Z}_{k}} \\ 
&\approx(2\pi)^{F}\ket{\mathcal{Z}_{k}(2\epsilon)}\bra{\mathcal{Z}_{k}}\delta(\mathcal{Z}_{k+1}-\mathcal{Z}_{k}(\epsilon)) .
\end{split}
\end{equation}
The standard PLDM/FBTS correlation function is then obtained by inserting the resulting approximate expressions for the electronic transition amplitudes into Eq.~(\ref{eq:corr_lin}) and then performing the integrals over the nuclear difference coordinates, $\Delta x_{k}$ and $\Delta p_{k}$, analytically:\cite{Huo2011densitymatrix,Hsieh2012FBTS}
\begin{equation}
\label{eq:corr_lin4}
\begin{split}
C_{AB}(t)\approx\int&\rd x\,\rd p\,\rd\mathcal{Z}\,\rd\mathcal{Z}' \\
\times\rho_{\text{b}}(x,p)&\phi(\mathcal{Z})\phi(\mathcal{Z}')A_{\text{m}}(\mathcal{Z},\mathcal{Z}')B_{\text{m}}(\mathcal{Z}'(t),\mathcal{Z}(t)) .
\end{split}
\end{equation}
In this expression, $\phi(\mathcal{Z})$ is the normalized Gaussian distribution used to sample the Cartesian mapping variables:
\begin{equation}
\label{eq:phi}
\phi(\mathcal{Z})=\frac{\eu{-\tfrac{1}{2}|\mathcal{Z}|^{2}}}{(2\pi)^{F}}
\end{equation}
and the electronic operators within the real-time quantum correlation function are now described in standard PLDM in terms of these mapping variables as:
\begin{equation}
\label{eq:coherent_op}
A_{\text{m}}(\mathcal{Z},\mathcal{Z}')=\frac{1}{2}\sum_{\lambda,\lambda'}\braket{\lambda|\hat{A}|\lambda'}Z_{\lambda}^{*}Z'_{\lambda'} .
\end{equation}
The dynamics of the mapping coherent states and the dynamics of the nuclear phase-space variables can be described by evolving the classical nuclear and electronic degrees of freedom under the following equations of motion:\cite{Huo2011densitymatrix,Hsieh2012FBTS}
\begin{equation}
\label{eq:harm-map_evolve}
\begin{split}
&\frac{\rd Z_{\lambda}}{\rd t}=-i\sum_{\mu}\braket{\lambda|\hat{V}(x)|\mu}Z_{\mu}, \\
&\frac{\rd Z'_{\lambda}}{\rd t}=-i\sum_{\mu}\braket{\lambda|\hat{V}(x)|\mu}Z'_{\mu}, \\
&\frac{\rd x}{\rd t}=\frac{p}{m},\qquad\frac{\rd p}{\rd t}=F(\mathcal{Z},\mathcal{Z}',x) = F_{0}(x)+F_{\text{e}}(\mathcal{Z},\mathcal{Z}',x) ,
\end{split}
\end{equation}
where $F_{0}(x)$ is the state-independent nuclear force, given by Eq.~(\ref{eq:force_indep}) and $F_{\text{e}}(\mathcal{Z},\mathcal{Z}',x)$ is the state-dependent nuclear force:
\begin{equation}
\label{eq:force_map}
F_{\text{e}}(\mathcal{Z},\mathcal{Z}',x)=-\tfrac{1}{2}\left[\nabla V_{\text{m}}(\mathcal{Z},x)+\nabla V_{\text{m}}(\mathcal{Z}',x)\right].
\end{equation}

\begin{figure}
\resizebox{0.4\textwidth}{!}{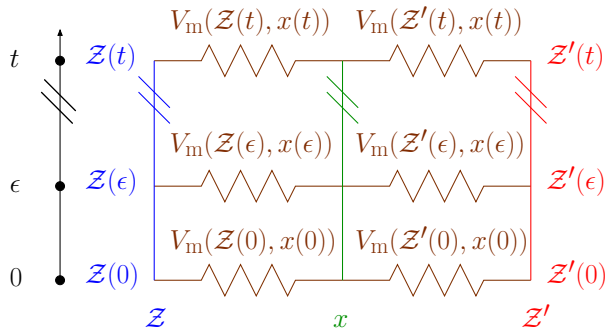}
\caption{Schematic illustrating the structure of the PLDM and spin-PLDM equations of motion, given by Eq.~(\ref{eq:harm-map_evolve}). The electronic mapping variables for the forward and backward paths ($\mathcal{Z}$ and $\mathcal{Z}'$ respectively) are not directly coupled, but instead couple via the nuclear degrees of freedom, $x$, through the interaction, $V_{\text{m}}$.}\label{fig:pldm}
\end{figure}

The discretized PLDM equations of motion are illustrated schematically in Fig.~\ref{fig:pldm}, for time-steps of size $\epsilon$. The electronic mapping variables for the forward and backward paths ($\mathcal{Z}$ and $\mathcal{Z}'$ respectively) are not directly coupled, but instead couple via the nuclear degrees of freedom, $x$, through $F_{\text{e}}(\mathcal{Z},\mathcal{Z}',x)$. This coupling within the equations of motion means that the back action of the electronic subsystem on the nuclear environment is included, at least within the approximations of the method.

The equations of motion for the standard PLDM approach previously derived by Huo and Coker\cite{Huo2011densitymatrix} differ slightly from the expression in Eq.~(\ref{eq:harm-map_evolve}). This is because within the standard PLDM approach, the potential is not separated into a purely nuclear potential, $V_{0}(x)$, and an electronic-dependent potential, $\hat{V}(x)$. This results in the whole nuclear force explicitly depending on the mapping variables $\mathcal{Z}$ and $\mathcal{Z}'$. The forward-backward trajectory solution (FBTS),\cite{Hsieh2012FBTS,Hsieh2013FBTS} which does separate the nuclear potential into $V_{0}(x)$ and $\hat{V}(x)$ components, has identical equations of motion as Eq.~(\ref{eq:harm-map_evolve}) and also has the same expression for the real-time correlation function as Eq.~(\ref{eq:corr_lin4}). FBTS is however derived as an approximate solution to the QCLE\cite{Kapral1999} and not from a path-integral representation of the real-time correlation function. Within the spin-mapping approach which we will introduce in Sec.~\ref{sec:spin-pldm}, the separation of these two nuclear potential components arises naturally from the underlying theory and hence it is no longer an arbitrary choice that one must make during the derivation of mapping-based methods.     

As the QCLE is derived by taking a partial Wigner transform of the exact time-dependent Schr{\"o}dinger equation, it describes the exact dynamics of electron-nuclear systems in the classical nuclear limit. In order to get a single trajectory picture of the electronic transition amplitudes, the derivation of both standard PLDM and FBTS apply the single approximation given by Eq.~(\ref{eq:coherent_delta}) for the overlap of harmonic oscillator coherent states at time-steps $k$ and $k+1$ to an otherwise exact mapping variable solution of the QCLE\@. Hence, in principle a more accurate solution of the QCLE dynamics could be found by improving upon the approximation given by Eq.~(\ref{eq:coherent_delta}). To do so, we derive spin-PLDM within the spin-mapping space.
\subsection{\label{sec:spin-pldm}Spin-PLDM}
The main idea behind spin-mapping is the representation of the electronic states of a system by a set of spins, described by classical vectors with fixed radii. This can be achieved in several ways, leading to different properties of the resulting dynamics. The most analogous form of spin-mapping to that of MMST mapping is to represent a $F$ electronic state system by $F$ spin-$\tfrac{1}{2}$ particles.\cite{Cotton2015spin} Like MMST mapping, this spin-mapping is mathematically described as a set of $F$ independent degrees of freedom, except each degree of freedom can now only contain 0 or 1 excitation, corresponding to the two quantum states of each spin-$\tfrac{1}{2}$ particle. While the size of the unphysical subspace available to this spin-mapping is clearly smaller than that of the MMST mapping space, it is still larger than necessary, and more importantly it also has the unfavourable feature that the classical dynamics are not exact for the pure electronic subsystem.\cite{Liu2016} This probably explains why this particular form of spin-mapping has been found to give rise to less accurate dynamical observables compared to MMST mapping for many systems.\cite{Cotton2015spin}

A seemingly more successful version of spin-mapping involves parameterizing the probability amplitudes associated with the electronic basis in terms of spin angles. Now the associated classical dynamics, which take the form of spins precessing around an effective magnetic field, exactly describe the pure electronic subsystem dynamics and the spin-mapping space now contains no unphysical regions. While this form of spin-mapping has been around since the advent of mapping-based techniques,\cite{Meyer1979spinmatrix,Meyer1979nonadiabatic} the underlying equations have recently been reformulated using a Stratonovich--Weyl approach, which leads to a more rigorous way of determining the optimum spin radius (or equivalently the zero-point energy parameter),\cite{spinmap} different expressions for the measurement operators relative to MMST mapping, as well as a generalization to systems with more than two levels. \cite{multispin} This reformulation has already been shown to produce more accurate observables compared to MMST mapping when applied to fully linearized techniques. The multi-state generalization of this spin mapping was formulated in terms of Cartesian mapping variables instead of spin angles and hence it is this approach that we will use to derive a spin-mapping version of PLDM.

Within this reformulation of spin-mapping, it was also shown that the theory can be equally implemented on many different spin spheres. For the fully linearized spin methods,\cite{spinmap,multispin} it was found that the W-sphere consistently gave the most accurate results. This is perhaps because the Stratonovich--Weyl W-functions are self-dual\cite{spinmap} and is also consistent with previous work that found that using the zero-point energy parameter for the W-sphere is in some sense optimum for MMST classical trajectories.\cite{Mueller1999pyrazine,Golosov2001,Cotton2013mapping,Cotton2016SQC} Preliminary tests performed by us have also shown that PLDM is systematically more accurate on the W-sphere than on the Q- and P-spheres. Hence we will only consider PLDM on the W-sphere, which we will refer to as spin-PLDM\@.

The Stratonovich--Weyl transform\cite{stratonovich1957} can be thought of as a discrete version of the Wigner transform within the spin-mapping space. Within this formalism, operators are represented by their Stratonovich--Weyl W-functions, which are given as follows in terms of the Cartesian mapping variables, $\mathcal{Z}$:
\begin{equation}
\label{eq:stratonovich_function}
A_{\text{W}}(\mathcal{Z})=\text{tr}[\hat{A}\hat{w}_{\text{W}}(\mathcal{Z})] ,
\end{equation}
where $\hat{w}_{\text{W}}(\mathcal{Z})$ is the Stratonovich--Weyl W-kernel, which is defined as:
\begin{equation}
\label{eq:stratonovich_kernel}
\braket{\mu|\hat{w}_{\text{W}}(\mathcal{Z})|\lambda}=\tfrac{1}{2}\left(Z_{\mu}Z^{*}_{\lambda}-\gamma_{\text{W}}\delta_{\mu\lambda}\right) ,
\end{equation}
and $\gamma_{\text{W}}$ is the zero-point energy parameter for the W-sphere:\cite{multispin}
\begin{equation}
\label{eq:gamma_w}
\gamma_{\text{W}}=\frac{1}{F}(R_{\text{W}}^{2}-2) .
\end{equation}
Like for the Wigner transform, the trace of a product of two operators can be exactly written as an integral over the product of their corresponding Stratonovich--Weyl W-functions: $\text{tr}[\hat{A}\hat{B}]=\int\rd\mathcal{Z}\,\rho_{\text{W}}(\mathcal{Z})A_{\text{W}}(\mathcal{Z})B_{\text{W}}(\mathcal{Z})$. Note that only the expression for the zero-point energy parameter given by Eq.~(\ref{eq:gamma_w}) results in this property of the Stratonovich--Weyl W-functions being satisfied. Within this relation, $A_{\text{W}}(\mathcal{Z})$ and $B_{\text{W}}(\mathcal{Z})$ are the W-functions of their associated electronic operator, given by Eq.~(\ref{eq:stratonovich_function}) and $\rho_{\text{W}}(\mathcal{Z})$ is a distribution which confines the mapping variables to lie on the surface of the W-sphere of radius $R_{\text{W}}$:
\begin{equation}
\label{eq:hypersphere}
\rho_{\text{W}}(\mathcal{Z}) = F\frac{\delta(|\mathcal{Z}|^{2}-R_{\text{W}}^{2})}{\int\rd\mathcal{Z}\,\delta(|\mathcal{Z}|^{2}-R_{\text{W}}^{2})},
\end{equation}
where the W-sphere radius, $R_{\text{W}}$, is given generally for a $F$-level electronic system as:\cite{Tilma2011,multispin}
\begin{equation}
\label{eq:wsphere_radius}
R_{\text{W}}^{2}=2\sqrt{F+1} .
\end{equation}
Additionally, a factor of $F$ is included in this definition of the Cartesian mapping variable distribution [Eq.~(\ref{eq:hypersphere})] so that $\int\rd\mathcal{Z}\rho_{\text{W}}(\mathcal{Z})=\text{tr}[\hat{\mathcal{I}}]=F$.

The Stratonovich--Weyl W-kernel can also be used to represent electronic operators; namely:
\begin{subequations}
\label{eq:kernel_properties}
\begin{align}
\label{eq:kernel_properties_ident}
&\hat{\mathcal{I}}=\int\rd\mathcal{Z}\,\rho_{\text{W}}(\mathcal{Z})\hat{w}_{\text{W}}(\mathcal{Z}) , \\ \label{eq:kernel_properties_op}
&\hat{V}(x)=\int\rd\mathcal{Z}\,\rho_{\text{W}}(\mathcal{Z})\hat{w}_{\text{W}}(\mathcal{Z})V_{\text{m}}(\mathcal{Z},x) ,
\end{align}
\end{subequations}
where $V_{\text{m}}(\mathcal{Z},x)$ is the MMST mapping representation of the traceless electronic Hamiltoninan $\hat{V}(x)$ [Eq.~(\ref{eq:map_ham})], which can be shown to be identical to the Stratonovich--Weyl W-function for the same operator from Eqs.~(\ref{eq:stratonovich_function}) and (\ref{eq:stratonovich_kernel}). Hence the coupling propagator, which appears in the definition of the electronic transition amplitudes, can also therefore be obtained in terms of the Stratonovich--Weyl W-kernel as:
\begin{equation}
\label{eq:spin_force_kernel}
\begin{split}
&\text{e}^{-\tfrac{i}{2}\nabla\hat{V}(x_{k})\epsilon \Delta x_{k}}= \\
&\int\rd\mathcal{Z}_{k}\,\rho_{\text{W}}(\mathcal{Z}_{k})\eu{-\tfrac{i}{2}\nabla V_{\text{m}}(\mathcal{Z}_{k},x_{k})\epsilon \Delta x_{k}}\hat{w}_{\text{W}}(\mathcal{Z}_{k})+\mathcal{O}(\epsilon^{3}) . 
\end{split}
\end{equation}
This result can be confirmed by comparing a Taylor expansion of the exponential on both sides of Eq.~(\ref{eq:spin_force_kernel}) and using the properties of the Stratonovich--Weyl kernel given by Eq.~(\ref{eq:kernel_properties}). As the error in this expression is of order $\epsilon^{3}$,\footnote{For other spin spheres, such as the Q- or P-sphere, the error associated with representing the coupling propagator in terms of the associated Stratonovich--Weyl kernel is of order $\epsilon^{2}$} the Stratonovich--Weyl representation of the coupling propagator is hence valid when the $N\rightarrow\infty$ limit is taken. This step is analogous to representing the coupling operator in terms of harmonic oscillator coherent states within the derivation of standard PLDM/FBTS [Eq.~(\ref{eq:mmst_force_kernel})].

Inserting Eq.~(\ref{eq:spin_force_kernel}) into Eq.~(\ref{eq:trans_amp_lina}) results in a Stratonovich--Weyl kernel appearing to the right of every electronic propagator (except for the $\eu{-i\hat{V}(x_{1})\epsilon}$ operator, which we post-multiply by $\hat{\mathcal{I}} = \int \rd \mathcal{Z}_{0}\,\rho_{\text{W}}(\mathcal{Z}_{0})\hat{w}_{\text{W}}(\mathcal{Z}_{0})$). We then define the time-evolved Stratonovich--Weyl kernel, $\hat{w}_{\text{W}}(\mathcal{Z}_{k},\epsilon)$, as follows: 
\begin{equation}
\label{eq:kernel_evolve}
\hat{w}_{\text{W}}(\mathcal{Z}_{k},\epsilon)=\eu{-i\hat{V}(x_{k+1})\epsilon}\hat{w}_{\text{W}}(\mathcal{Z}_{k}) .
\end{equation}
Unlike for the harmonic oscillator coherent states used in standard PLDM, the time-evolved Stratonovich--Weyl kernel cannot be obtained purely
by evolving the Cartesian mapping variables, $\mathcal{Z}_{k}$. This is because the Stratonovich--Weyl kernel contains a zero-point energy parameter, $\gamma_{\text{W}}$, multiplied by an identity operator which must also be evolved forward in time. Hence, the time-evolved Stratonovich--Weyl kernel is given by:
\begin{equation}
\begin{split}
&\braket{\mu|\hat{w}_{\text{W}}(\mathcal{Z}_{k},\epsilon)|\lambda} \\
&=\tfrac{1}{2}\left(Z^{(k)}_{\mu}(\epsilon)\left[Z^{(k)}_{\lambda}\right]^{*}-\gamma_{\text{W}}\braket{\mu|\eu{-i\hat{V}(x_{k+1})\epsilon}|\lambda}\right) ,
\end{split}
\end{equation}
where $\mathcal{Z}_{k}(\epsilon)=\{Z_{1}^{(k)}(\epsilon),Z_{2}^{(k)}(\epsilon),\cdots,Z_{F}^{(k)}(\epsilon)\}$ are the Cartesian mapping variables at time $\epsilon$, evolved according to Eq.~(\ref{eq:harm-map_evolve}).
These equations of motion for the Cartesian mapping variables exactly describe the pure electronic sub-system dynamics, for any size time-step $\epsilon$. Additionally, $\eu{-i\hat{V}(x_{k+1})\epsilon}$ is an $F\times F$ matrix in the electronic-state basis, which can easily be computed numerically. For the backward ($\mathcal{Z}'$) electronic transition amplitude, inserting Eq.~(\ref{eq:spin_force_kernel}) into Eq.~(\ref{eq:trans_amp_linb}) results in a Stratonovich--Weyl kernel appearing to the left of every electronic propagator, $\eu{i\hat{V}(x_{k})\epsilon}$. Hence the Hermitian conjugate of Eq.~(\ref{eq:kernel_evolve}) is used in this case.

Therefore, Eqs.~(\ref{eq:spin_force_kernel}) and (\ref{eq:kernel_evolve}) lead to an expression for the electronic transition amplitudes in terms of the Cartesian mapping variables. However, the expression contains an integral over mapping variables at each time-step, $k$. Following the standard PLDM/FBTS procedure, in analogy to Eq.~(\ref{eq:coherent_delta}), the overlap of Stratonovich--Weyl kernels at different time-steps are approximated as follows, to generate a trajectory picture for the electronic dynamics:
\begin{equation}
\label{eq:spin_kernel_delta}
\hat{w}_{\text{W}}(\mathcal{Z}_{k+1},\epsilon)\hat{w}_{\text{W}}(\mathcal{Z}_{k},\epsilon)\approx\hat{w}_{\text{W}}(\mathcal{Z}_{k},2\epsilon)\delta(\mathcal{Z}_{k+1}-\mathcal{Z}_{k}(\epsilon)) .
\end{equation}
where the Dirac delta function of a complex argument, $\mathcal{Z}$, is defined as: $\delta(\mathcal{Z})=\delta(\text{Re}[\mathcal{Z}])\delta(\text{Im}[\mathcal{Z}])$. Employing the approximation given by Eq.~(\ref{eq:spin_kernel_delta}) reduces the expression for the electronic transition amplitudes to:
\begin{subequations} \label{eq:trans_amp_lin_kernel}
\begin{align}
&T_{[\mu,\lambda]}\approx\int\rd\mathcal{Z}\rho_{\text{W}}(\mathcal{Z})\braket{\mu|\hat{w}_{\text{W}}(\mathcal{Z},t)|\lambda}\eu{iS_{\text{e}}} , \\
&T'_{[\lambda',\mu']}\approx\int\rd\mathcal{Z}'\rho_{\text{W}}(\mathcal{Z}')\braket{\lambda'|\hat{w}^{\dagger}_{\text{W}}(\mathcal{Z}',t)|\mu'}\eu{-iS'_{\text{e}}} ,
\end{align}
\end{subequations}
which is exact for a purely electronic system and where we now set $\mathcal{Z}_{0}=\mathcal{Z}$ and $\mathcal{Z}'_{0}=\mathcal{Z}'$. The electronic action for the forward and backward paths are defined as:
\begin{subequations} \label{eq:kernel_elec_action}
\begin{align}
&S_{\text{e}}=-\frac{1}{2}\sum_{k=1}^{N}\nabla V_{\text{m}}(\mathcal{Z}(t_{k}),x_{k})\epsilon\Delta x_{k} , \\ 
&S'_{\text{e}}=\frac{1}{2}\sum_{k=1}^{N}\nabla V_{\text{m}}(\mathcal{Z}'(t_{k}),x_{k})\epsilon\Delta x_{k} ,
\end{align}
\end{subequations}
where $t_{k}=k\epsilon$ is the time at time-step $k$. Additionally, $\hat{w}_{\text{W}}(\mathcal{Z},t)$ is the time-evolved Stratonovich--Weyl kernel, which is defined as:
\begin{equation}
\label{eq:kernel_prop}
\braket{\mu|\hat{w}_{\text{W}}(\mathcal{Z},t)|\lambda}=\tfrac{1}{2}\left(Z_{\mu}(t)Z^{*}_{\lambda}-\gamma_{\text{W}}\braket{\mu|\hat{U}(t)|\lambda}\right) ,
\end{equation}
where $\hat{U}(t)$ is is a time-ordered propagator for the electronic states according to the time-dependent potential obtained by moving along a given path $x(t_k)=x_{k}$:
\begin{equation}
\label{eq:time-ordered-prop}
\hat{U}(t)=\eu{-i\hat{V}(x_{N})\epsilon}\cdots\eu{-i\hat{V}(x_{2})\epsilon}\eu{-i\hat{V}(x_{1})\epsilon} .
\end{equation}
Assuming that $F$ is not too large, $\hat{U}(t)$ is easy to evaluate as it is just the dynamics of the bare electronic space according to a time-dependent Hamiltonian. The back action does not appear directly in the time-ordered propagator, but is treated by the evolution of the Cartesian mapping variables.

To complete the linearization approximation of the nuclear path, outlined in Sec.~\ref{sec:theory}, the integrals over $\Delta x_{k}$ and $\Delta p_{k}$ must be performed. Using the approximate expressions for the nuclear and electronic action, given by Eqs.~(\ref{eq:nuc_action_lin}) and (\ref{eq:kernel_elec_action}), these integrals can be performed analytically in a similar fashion to the standard PLDM approach \cite{Huo2011densitymatrix} to give:
\begin{equation}
\begin{split}
\label{eq:lin_int}
&\int\frac{\rd\Delta p_{0}}{(2\pi)^{f}}\prod_{k=1}^{N-1}\frac{\rd\Delta x_{k}}{(2\pi)^{f}}\frac{\rd\Delta p_{k}}{(2\pi)^{f}}\eu{i(S_{\text{e}}-S'_{\text{e}})}\eu{i\Delta S_{0}} \\
&\approx\prod_{k=0}^{N-1}\delta\left(\frac{x_{k+1}-x_{k}}{\epsilon}-\frac{p_{k}}{m}\right)\prod_{k=1}^{N-1}\delta\left(\frac{p_{k}-p_{k-1}}{\epsilon}-F_{k}\right) .
\end{split}
\end{equation}
The result is a product of Dirac delta functions which constrain the dynamics of the nuclear variables to follow Newton's equation of motion. The equations of motion for the classical variables within spin-PLDM are identical to those of standard PLDM [Eq.~(\ref{eq:harm-map_evolve})], although the correlation functions will of course still give different results due to the change in the initial distribution of the Cartesian mapping variables.

This means that the approximate expression for the spin-PLDM real-time correlation function becomes:
\begin{equation}
\label{eq:corr_lin_cartesian}
\begin{split}
C_{AB}(t)&\approx\int\rd x\,\rd p\,\rd\mathcal{Z}\,\rd\mathcal{Z}'\,\rho_{\text{W}}(\mathcal{Z})\rho_{\text{W}}(\mathcal{Z}')\rho_{\text{b}}(x,p) \\
&\times\tr\left[\hat{A}\hat{w}^{\dagger}_{\text{W}}(\mathcal{Z}',t)\hat{B}\hat{w}_{\text{W}}(\mathcal{Z},t)\right] .
\end{split}
\end{equation}
where we now set $x_{0}=x$ and $p_{0}=p$. The trace in Eq.~(\ref{eq:corr_lin_cartesian}) only contains $F$ terms and is hence easy to perform explicitly for the majority of systems of interest. 

The spin-PLDM algorithm can be easily implemented as follows. First, the integrals contained in the correlation function given by Eq.~(\ref{eq:corr_lin_cartesian}) can be evaluated numerically by sampling the initial classical coordinates using Monte Carlo. This means sampling the nuclear phase-space variables from the Wigner distribution of the nuclear initial density [Eq.~(\ref{eq:wigner})] and sampling the Cartesian mapping variables for the forward and backward paths ($\mathcal{Z}$ and $\mathcal{Z}'$) independently from uniform hyperspheres of radius $R_{\text{W}}$. For each instance of the sampling, the classical coordinates can then be evolved in time using the equations of motion given by Eq.~(\ref{eq:harm-map_evolve}) and additionally the time-ordered propagator can be obtained by using Eq.~(\ref{eq:time-ordered-prop}), where $\epsilon$ is the time-step. Finally, the W-kernel for both the forward and backward paths can then be constructed using Eq.~(\ref{eq:kernel_prop}) and the contribution to the real-time correlation function for each sample can be obtained by explicitly evaluating the matrix multiplications and trace in the electronic basis. 
Such an algorithm can also be implemented with only a few minor changes into a preexisting standard PLDM code. 

The main differences between spin-PLDM and standard PLDM (whose correlation function is given by Eq.~(\ref{eq:corr_lin4})) are as follows. In spin-PLDM, the initial mapping variables are constrained to a hypersphere $|\mathcal{Z}|^{2}=R^{2}_{\text{W}}$, which guarantees that the electronic state being represented is correctly normalized. Additionally, spin-PLDM has a zero-point energy parameter, which means that it treats correlation functions containing the identity operator slightly differently from those of traceless operators. These two changes are expected to lead to improved results in a similar way in which 
fully linearized spin-mapping\cite{spinmap,multispin}  demonstrates improved accuracy over fully linearized MMST mapping-based approaches.
\section{Results}
In this section, we consider a set of challenging systems far from the Born--Oppenheimer limit. In order to improve the convergence of the spin-PLDM method with respect to the number of trajectories, we have obtained the following results using so-called focused initial conditions to sample the Cartesian mapping variables at $t=0$. The details concerning the implementation of these focused initial conditions for spin-PLDM can be found in Paper II\cite{paper2} along with numerical comparisons of the two approaches. In contrast to MMST mapping based methods, we have found that the spin-PLDM results obtained using focused initial conditions are essentially indistinguishable from those obtained using the original sampling of the Cartesian mapping variables, as outlined in Sec.~\ref{sec:spin-pldm}, although the later require more trajectories. For the MMST mapping-based techniques, only the most accurate form of the methods are used, so as to act as a fair comparison with spin-mapping; hence we present results only for the non-focused variants of these methods. 
\subsection{The spin-boson model}
To test the accuracy of our spin-PLDM method, we have applied the method to a series of spin-boson models. The parameters associated with these spin-boson models are chosen in order to cover a large part of the parameter space, including both symmetric and asymmetric systems, low and high temperature limits and strong and weak system-bath coupling. Such spin-boson models are commonly used to benchmark new methods, as numerically exact results are available for comparison. The model essentially consists of two electronic states coupled to an initially thermalized harmonic bath, whose Hamiltonian is given by:\cite{Garg1985spinboson}
\begin{subequations}
\begin{align}
\label{eq:spinboson_ham}
\hat{H}&=\tfrac{1}{2}\sum_{j=1}^{f}\hat{p}_{j}^{2} + V_{0}(\hat{x}) + \hat{V}(\hat{x}), \\
V_{0}(\hat{x})&=\tfrac{1}{2}\sum_{j=1}^{f}\omega_{j}^{2}\hat{x}_{j}^{2}, \\
\hat{V}(\hat{x})&=\Delta\hat{\sigma}_{x}+\left(\varepsilon+\sum_{j=1}^{f}c_{j}\hat{x}_{j}\right)\hat{\sigma}_{z} .
\end{align}
\end{subequations}
Here, $\varepsilon$ is the energy bias and $\Delta$ is the constant diabatic coupling. The bath contains $f$ nuclear modes, each with frequency $\omega_{j}$ and electron-nuclear coupling coefficient $c_{j}$. The mass, $m_{j}$, has been incorporated into the definition of the nuclear position, $\hat{x}_{j}$, and momentum, $\hat{p}_{j}$, operators. As the nuclear bath is harmonic, the dynamics of the spin-boson model is exactly described by the quantum-classical Liouville equation (QCLE).\cite{MacKernan2002,Shi2004QCLE} Hence any errors arising from calculating correlation functions using mapping-based classical-trajectory techniques arise solely from approximations to the QCLE and are not inherently due to the classical treatment of the nuclear degrees of freedom. The spectral density, $J_{\text{bath}}(\omega)$, determines the distribution of nuclear frequencies within the bath. One of the most commonly used spectral densities is the Ohmic bath form:
\begin{equation}
\label{eq:bath_spectral}
J_{\text{bath}}(\omega)=\frac{\pi\xi}{2}\omega\,\eu{-\omega/\omega_{\text{c}}} ,
\end{equation}
where $\omega_{\text{c}}$ is the characteristic frequency and $\xi$ is the Kondo parameter. In order to perform numerical simulations, the continuous bath must first be discretized into a finite number of modes. The spectral density can then be represented as follows:
\begin{equation}
\label{eq:bath_discrete}
J_{\text{bath}}(\omega)=\frac{\pi}{2}\sum_{j=1}^{f}\frac{c_{j}^{2}}{\omega_{j}}\delta(\omega-\omega_{j}) .
\end{equation}
We have used the discretization scheme employed in Ref.~\onlinecite{Craig2005}. Additionally, we consider solely correlation functions in which the uncoupled bath is initially thermalized. This means that for mapping-based classical-trajectory techniques, the initial nuclear coordinates are sampled from the following Wigner distribution:
\begin{equation}
\label{eq:bath_wigner}
\rho_{\text{b}}(x,p)=\prod_{j=1}^{f}\frac{\alpha_{j}}{\pi}\text{exp}\left[-\frac{2\alpha_{j}}{\omega_{j}}\left(\tfrac{1}{2}p_{j}^{2}+\tfrac{1}{2}\omega_{j}^{2}x_{j}^{2}\right)\right] ,
\end{equation}
where $\alpha_{j}=\text{tanh}(\tfrac{1}{2}\beta\omega_{j})$ and $\beta$ is the inverse temperature. For comparison, numerically exact results for the real-time quantum correlation functions have been obtained using the quasiadiabatic path-integral (QUAPI) technique.\cite{Makarov1994}

The parameters for the spin-boson models we consider are given by rows (a)--(f) in Table \ref{tab:spinboson_param}. For these calculations, we have used $f=100$ nuclear degrees of freedom, except for the strong coupling systems (e) and (f), where $f=400$ nuclear degrees of freedom were needed for convergence. Such a set of spin-boson models cover a wide range of different physical regimes, from symmetric to asymmetric, weak to strong system-bath coupling and low to high temperature limits. Hence these models offer a comprehensive test for our new spin-PLDM method. Additionally, all of these spin-boson models lie in a challenging regime for classical-trajectory methods to describe correctly, as they are far from the Born--Oppenheimer limit and have already been used to test other methods, which allows us to directly compare the accuracy of various methods.\cite{Cotton2016SQC,spinmap,Tao2016,linearized} While Ehrenfest and standard linearized mapping methods are known to perform relatively well for the symmetric models, these methods fail to correctly predict the long-time populations in the asymmetric models, due to a failure to obey detailed balance. As in Ref.~\onlinecite{linearized}, we also make a point of computing the time-dependent coherences, as well as populations. In this paper, Eq.~(\ref{eq:bath_wigner}) initializes the nuclear coordinates from the Boltzmann distribution of nuclear potential $V_{0}(x)$. This is subtly different from Refs.~\onlinecite{Cotton2016SQC} and \onlinecite{Tao2016}, where the nuclei are sampled from the Boltzmann distribution for potential $V_{0}(x)+\braket{1|\hat{V}(x)|1}$. Nonetheless, these different bath initial conditions do not appear to lead to significant differences in the results. 
\begin{table}
    \centering
    \caption{The spin-boson model parameters corresponding to the panels in Fig~\ref{fig:spin-boson}.
    The energy scale is defined in units of $\Delta$.}
    \label{tab:spinboson_param}
    \begin{ruledtabular}
    \begin{tabular}{ccccc}
        Model  &  $\varepsilon/\Delta$ & $\xi$ & $\beta\Delta$ & $\omega_{\text{c}}/\Delta$ \\  \hline
        (a) & 0 & 0.09 & 0.1 & 2.5 \\
        (b) & 0 & 0.09 & 5 & 2.5 \\
        (c) & 1 & 0.1 & 0.25 & 1 \\
        (d) & 1 & 0.1 & 5 & 2.5 \\
        (e) & 0 & 2 & 1 & 1 \\
        (f) & 5 & 4 & 0.1 & 2 \\
    \end{tabular}
    \end{ruledtabular}
\end{table}

\begin{figure*}
\includegraphics[scale=.90]{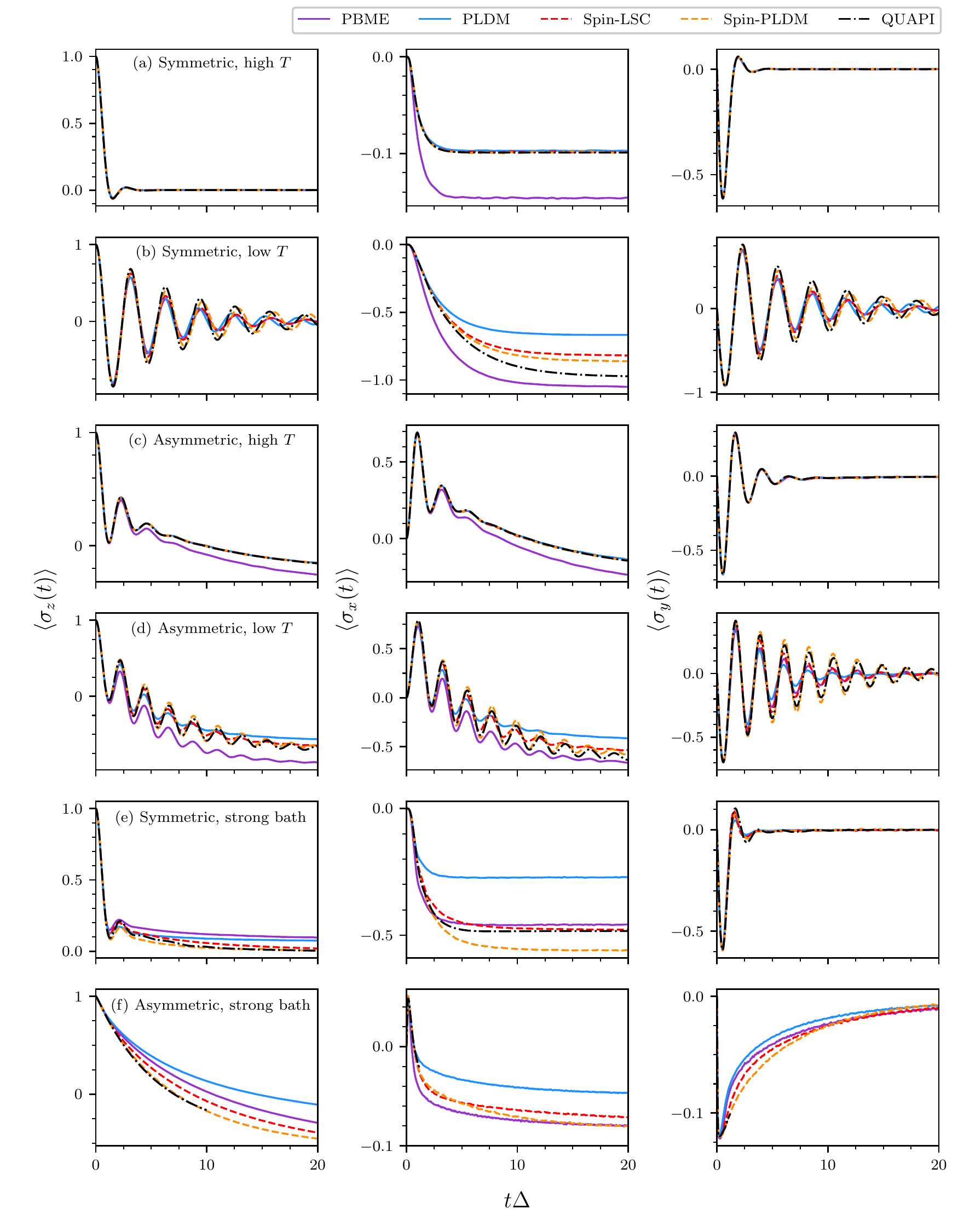}
\caption{Comparison of a range of mapping-based methods, for several different spin-boson models, with parameters defined in Table \ref{tab:spinboson_param}. For each system, the electronic system initially begins in the high energy diabatic state $\ket{1}$ and then the expectation values of the Pauli spin matrices, $\hat{\sigma}_{j}$, are recorded as a function of time. The dashed black lines give the numerically exact results, obtained using QUAPI. All results for spin-mapping based techniques (spin-LSC and spin-PLDM) correspond to their focused variant, while all results for MMST mapping based techniques (PBME and PLDM) correspond to their non-focused variant. The reader is encouraged to compare these results with those calculated using symmetrical quasi-classical (SQC) windowing (Figs.~6 and 7 from Ref.~\onlinecite{Cotton2016SQC} and Fig.~2 from Ref.~\onlinecite{linearized}), Ehrenfest (Figs.~3 from Ref.~\onlinecite{spinmap} and Fig.~2 from Ref.~\onlinecite{linearized}) and traceless MMST (Fig.~2 from Ref.~\onlinecite{linearized}). Our spin-LSC (Fig.~3 from Ref.~\onlinecite{spinmap}) and standard PLDM (Figs.~5(a), 5(c) and 5(d) from Ref.~\onlinecite{Mandal2018quasidiabatic}) results are in agreement with those already published.}\label{fig:spin-boson}
\end{figure*}

Fig.~\ref{fig:spin-boson} shows the calculated dynamical expectation values of all Pauli spin matrices for a range of mapping-based techniques. All results correspond to the system initially occupying the higher energy diabatic state, $\ket{1}$. The spin-LSC approach is the fully linearized version of spin-PLDM and corresponds to the method described in Ref.~\onlinecite{multispin} for the W-sphere. Additionally, the Poisson-bracket mapping equation (PBME) approach is the fully linearized version of standard PLDM\cite{Hsieh2012FBTS} and is described in Ref.~\onlinecite{Kim2008Liouville}. While we do not show results for LSC-IVR in this paper, it has been observed that both PBME and LSC-IVR produce real-time correlation functions to a similar degree of accuracy.\cite{identity,FMO,linearized} Consider first the weak system-bath coupling spin-boson models at high temperature (given by (a) and (c)). We would expect to describe the dynamics of these models correctly, as the classical approximation that is employed in deriving mapping-based classical-trajectory techniques should be valid at high temperatures. We find that only PBME is unable to accurately describe the dynamics within these spin-boson models. The error arises because PBME does not correctly calculate the identity-containing correlation functions. 
Note that this is not a fundamental problem of linearized MMST dynamics as it is known that the error can be almost completely removed
by using an `identity-trick' recently derived for PBME\cite{identity,FMO,linearized} or by windowing trajectories at time $t$ using SQC.\cite{Cotton2013SQC,Cotton2013mapping,Cotton2019} 
Therefore, for this parameter regime, spin-PLDM does not offer any advantage over spin-LSC or standard PLDM, which are essentially numerically exact already for these systems. 

Next, we consider the more difficult models with weak system-bath coupling at low temperatures (given by (b) and (d)). For these models, spin-PLDM appears to consistently produce the most accurate results compared to other mapping-based approaches. First, spin-PLDM outperforms spin-LSC for these systems, by correcting the overdamped coherences found in the spin-LSC correlation functions of traceless operators. Additionally, spin-PLDM also outperforms standard PLDM in this parameter regime, because restraining the Cartesian mapping variables to a hypersphere reduces the large errors observed in the identity-containing correlation functions. This in turn results in a reduction of the error associated with the long-time limit of any real-time correlation function, because all correlation functions containing traceless operators decay to zero as $t\rightarrow\infty$. The spin-PLDM initial conditions for the mapping variables also appears to correct for the overdamped coherences found in the standard PLDM correlation functions of traceless operators. While both SQC and the MMST identity-trick can to some extent correct for the errors in the long-time limits of the PBME correlation functions, they typically still predict overdamped coherences similarly to spin-LSC.\cite{linearized,Cotton2016SQC} 

Finally, systems (e) and (f) are the most challenging spin-boson models that we consider in this paper, due to the strong system-bath coupling. In fact, the coupling in system (f) is so strong that we struggled to converge the numerically exact QUAPI results. Hence we only present the QUAPI results for times for which we are certain that they are numerically exact. The main conclusion from these figures is that spin-PLDM is the best method at obtaining the short time dynamics correctly and therefore is also expected to be the most trustworthy method for predicting the long-time limit of dynamical quantities. We assume that any method which is less accurate than spin-PLDM at short times but appears more accurate in the long-time limit in isolated cases must be due to a lucky cancellation of errors. The figures show that spin-PLDM provides a larger correction to spin-LSC for the strong-coupling spin-boson models. This improvement even holds for the high-temperature model (f), where spin-LSC was already quite good. This suggests that the explicit treatment of the forward and backward electronic mapping variables in spin-PLDM are particularly important for describing the dynamics in systems with relatively strong system-bath coupling. 
\subsection{\label{sec:res_gen}The Fenna--Mathews--Olsen complex}
To demonstrate that the spin-PLDM method can be easily applied to multistate systems, we have applied the method to a $F=7$ model for the FMO complex, which is a commonly studied photosynthetic pigment-protein complex found in green sulfur bacteria.\cite{Fenna1975}  This is a challenging benchmark problem that has already been used to test a wide range of mapping-based techniques. \cite{multispin,FMO,Hsieh2013FBTS,Rhee2011,Kim2012,Cotton2019,Miller2016Faraday,Cotton2016_2,Huo2010,Huo2011densitymatrix,Huo2012MolPhys,Mandal2018quasidiabatic,Tao2010FMO,Pfalzgraff2019GQME,Teh2017,Rivera2013,Lee2016} However, as for the spin-boson model, the dynamics of this FMO model is exactly described by the QCLE, because the nuclear bath is harmonic, which means that nuclear quantum effects do not need to be included in order to describe the quantum dynamics of this model exactly.\cite{MacKernan2002,Shi2004QCLE,Kapral2015QCL} Numerically exact results have been computed for this model from the hierarchical equation of motion (HEOM) approach. \cite{Ishizaki2009,Ishizaki2009_2,Ishizaki2010,Zhu2011,Wilkins2015,Sarovar2010} 

The Hamiltonian of this model is:\cite{Ishizaki2009}
\begin{equation}
\label{eq:fmo_ham}
\hat{H}=\hat{H}_{\text{s}}+\hat{H}_{\text{b}}+\hat{H}_{\text{sb}} ,
\end{equation}
where the electronic system Hamiltonian is given, in units of $\text{cm}^{-1}$, as:\cite{Adolphs2006}
\begin{equation}
\label{eq:fmo_sys}
    \hat{H}_\text{s} = \begin{pmatrix} 
    12410 & -87.7 & 5.5 & -5.9 & 6.7 & -13.7 & -9.9 \\ 
    -87.7 & 12530 & 30.8 & 8.2 & 0.7 & 11.8 & 4.3 \\
    5.5 & 30.8 & 12210 & -53.5 & -2.2 & -9.6 & 6.0 \\
    -5.9 & 8.2 & -53.5 & 12320 & -70.7 & -17.0 & -63.3 \\
    6.7 & 0.7 & -2.2 & -70.7 & 12480 & 81.1 & -1.3 \\
    -13.7 & 11.8 & -9.6 & -17.0 & 81.1 & 12630 & 39.7 \\
    -9.9 & 4.3 & 6.0 & -63.3 & -1.3 & 39.7 & 12440
    \end{pmatrix}. \nonumber
\end{equation}
Each diabatic electronic state is coupled to its own independent harmonic bath that consists of $f$ nuclear modes with frequencies $\omega_{j}$. The mass-weighted Hamiltonian for all seven independent baths is therefore given by:
\begin{equation}
\label{eq:fmo_bath}
H_{\text{b}}(\hat{x})=\sum_{n=1}^{7}\sum_{j=1}^{f}\left(\tfrac{1}{2}\hat{p}_{j,n}^{2}+\tfrac{1}{2}\omega_{j}^{2}\hat{x}_{j,n}^{2}\right) ,
\end{equation}
where each bath has the same set of frequencies, $\omega_{j}$. The model then contains a linear coupling term, which connects each harmonic bath to its corresponding diabatic state as follows:
\begin{equation}
\label{eq:fmo_coup}
\hat{H}_{\text{sb}}(\hat{x})=\sum_{n=1}^{7}\sum_{j=1}^{f}c_{j}\hat{x}_{j,n}\ket{n}\bra{n} ,
\end{equation}
where $c_{j}$ are the exciton-nuclear coupling coefficients.

The spectral density, $J_{\text{bath}}(\omega)$, determines the distribution of nuclear frequencies within each of the baths and their couplings. This model employs the Debye spectral density:
\begin{equation}
\label{eq:bath_spectral_gen}
J_{\text{bath}}(\omega)=2\Lambda\frac{\omega\omega_{\text{c}}}{\omega^{2}+\omega^{2}_{\text{c}}} ,
\end{equation}
where $\omega_{\text{c}}$ is the characteristic frequency of the bath (with $\tau_{\text{c}}=\omega_{\text{c}}^{-1}$ its corresponding time scale) and $\Lambda$ is the reorganization energy. We use $\Lambda=35\,\text{cm}^{-1}$, to be consistent with previous work, and only consider the situation of a `fast bath' (i.e., $\tau_{\text{c}}=50$ fs), which is the most difficult previously studied parameter regime. In order to perform numerical simulations, the continuous bath must be first discretized into a finite number of modes, as shown in Eq.~(\ref{eq:bath_discrete}). We have used the discretization scheme employed in Ref.~\onlinecite{Craig2007condensed} with $f=60$ bath modes per state. We consider the dynamics after the initial excitation of a single site, with the baths in thermal equilibrium before the excitation. This means that the initial nuclear coordinates associated with each excitonic site are sampled from the same Wigner distribution given by Eq.~(\ref{eq:bath_wigner}). 

\begin{figure*}
\includegraphics{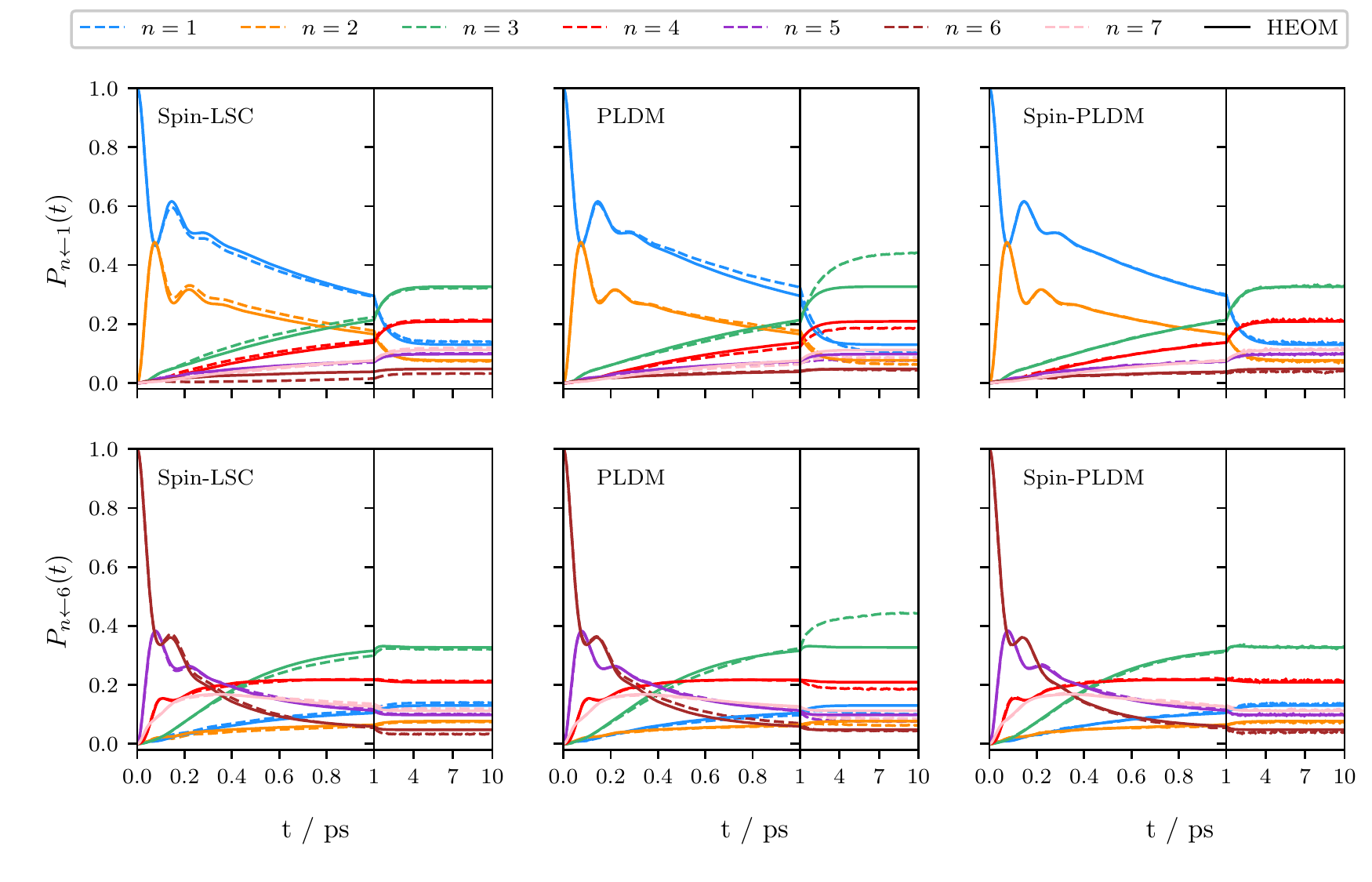}
\caption{The time-dependent diabatic populations for a seven-state FMO model at $T=300$K computed from various methods. For the first row of figures, the electronic system was started in diabatic state 1, whereas the second row of figures correspond to starting the system in diabatic state 6. Exact HEOM results\cite{Ishizaki2009,Wilkins2015} are given by the solid lines. All results for spin-mapping based techniques (spin-LSC and spin-PLDM) correspond to their focused variant, while the PLDM results correspond to the non-focused variant. We recommend that the reader compares these results with those calculated using symmetrical quasi-classical (SQC) windowing (Figs.~10(c) and 11(c) from Ref.~\onlinecite{Cotton2019}), Ehrenfest (Figs.~6 and S3 from Ref.~\onlinecite{multispin}) and traceless MMST (Fig.~4 from Ref.~\onlinecite{FMO}). Both our spin-LSC (Figs.~6 and S3 from Ref.~\onlinecite{multispin}) and standard PLDM (Fig. 6 from Ref.~\onlinecite{Huo2012MolPhys}) results are in agreement with those previously published.}\label{fig:fmo_300k}
\end{figure*}

To test the accuracy of our newly developed spin-PLDM method, we calculate the time-dependence of the exciton dynamics for this FMO model. Of particular interest is the picosecond exciton population transfer to the lowest energy site/chromophore ($n=3$), from which the excitons can be harvested at the reaction centre. We consider both the dynamics at high and low temperature (300\,K and 77\,K respectively) and also with the initial exciton residing on different sites/chromophores of the complex (i.e., site 1 and 6). Fig.~\ref{fig:fmo_300k} shows the population dynamics of the FMO at high temperature (300\,K).
As stated before, this is the regime in which the approximations involved in deriving mapping-based classical-trajectory techniques are expected to be valid. From Fig.~\ref{fig:fmo_300k}, it can be seen that spin-LSC, standard PLDM and spin-PLDM are each able to describe the short time dynamics very well (i.e., up to 1 ps), irrespective of whether the exciton is initialized on site 1 (the top row of figures) or site 6 (the bottom row of figures). Even for this short-time behaviour, the accuracy of the spin-PLDM approach however appears superior, as the calculated dynamics from this technique are essentially indistinguishable from the exact HEOM results (solid lines) for this high temperature parameter regime. The accuracy of spin-PLDM continues to be good, even when the long-time dynamics (up to 10 ps) are considered. The spin-LSC method also gives an excellent description of the long time limits of the exciton populations. However, standard PLDM is unable to describe the long time dynamics correctly for this high temperature regime and exhibits a large deviation from the exact result, in particular for the site 3 population. This illustrates a clear advantage in using spin-mapping instead of the standard MMST approach. Constraining the mapping variables to a hypersphere ensures that these Cartesian mapping variables remain in the physical subspace during the dynamics and this appears to minimize the errors produced in the long-time limit and also allows the method to better approximate the correct Boltzmann distribution. While both the SQC\cite{Cotton2019} and the MMST identity-trick\cite{FMO} results for this model are clearly superior to those for PBME,\cite{FMO} both these methods shower greater errors in the dynamical populations compared to spin-PLDM, even when just considering the evolution up to 1 ps.

\begin{figure*}
\includegraphics{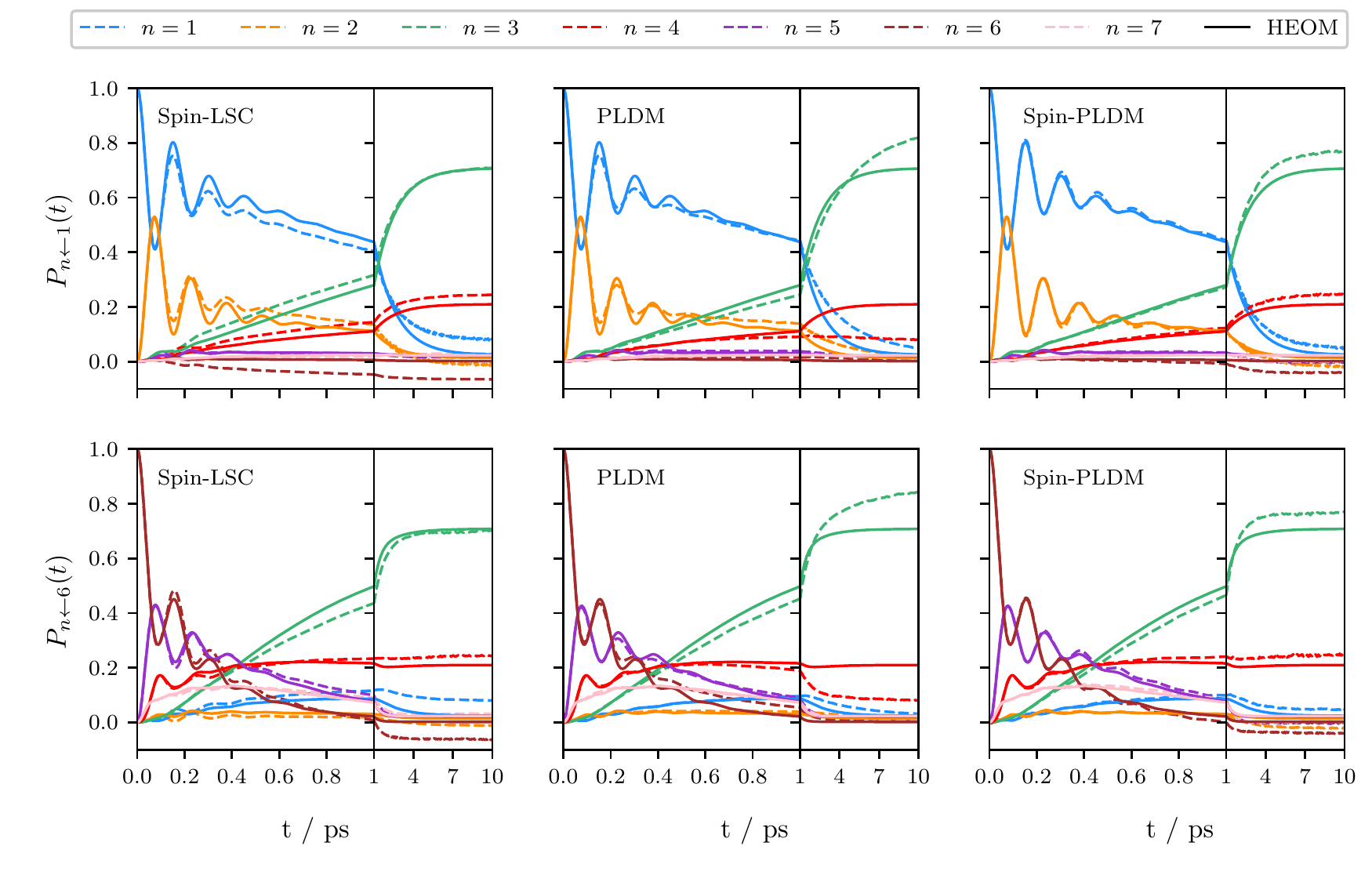}
\caption{The diabatic populations for a seven-state FMO model at $T=77$K. For the first row of figures, the electronic system was started in diabatic state 1, whereas the second row of figures correspond to starting the system in diabatic state 6. Exact HEOM results\cite{Ishizaki2009,Wilkins2015} are given by the solid lines. All results for spin-mapping based techniques (spin-LSC and spin-PLDM) correspond to their focused variant, while the PLDM results correspond to the non-focused variant. We recommend that the reader compares these results with those calculated using symmetrical quasi-classical (SQC) windowing (Figs.~10(a) and 11(a) from Ref.~\onlinecite{Cotton2019} and Figs.~3 and 4 of Ref.~\onlinecite{Cotton2016_2}), Ehrenfest (Figure 5 from Ref.~\onlinecite{Cotton2016_2} and Figs.~4, 5, S1 and S2 from Ref.~\onlinecite{multispin}) and traceless MMST (Figs.~2, 3 and 6 from Ref.~\onlinecite{FMO}). Both our spin-LSC (Figs.~4, 5, S1 and S2 from Ref.~\onlinecite{multispin}) and standard PLDM (Fig.~3 from Ref.~\onlinecite{Huo2011densitymatrix} and Fig.~6 from Ref.~\onlinecite{Mandal2018quasidiabatic}) results are in agreement with those previously published.}\label{fig:fmo_77k}
\end{figure*}

The more challenging regime corresponds to studying the dynamics of the model at a low temperature (77\,K). Fig.~\ref{fig:fmo_77k} compares the calculated population dynamics for the same mapping-based classical-trajectory techniques. 
In the short-time dynamics, spin-LSC now shows noticeable errors compared to the exact HEOM results. Additionally some of the calculated populations with spin-LSC now become negative, even at quite short times (see for example the $n=6$ population, when starting in site 1), which is unphysical. While the standard PLDM results appear fairly accurate at short times, the method suffers from overdamped coherences, similar to what was observed when applying the method to the spin-boson model. Spin-PLDM produces almost exact results for this short-time dynamics, which again illustrates its apparent increased accuracy, not only over spin-LSC and PLDM but also when compared with the MMST identity-trick \cite{FMO} and SQC. \cite{Cotton2019} In the long-time limit (i.e., up to 10 ps), the standard PLDM approach again leads to large errors in the populations. In contrast, spin-LSC appears to produce less severe errors in the populations in the long-time limit. In particular, the $n=3$ population is well reproduced by the method. This is however probably partly fortuitous, due to the method's inability to describe the short-time dynamics of this population correctly. The errors within spin-PLDM in the long-time limit also appear not too severe and the method also suffers less severely with negative populations compared to spin-LSC\@. Standard PLDM, in contrast, always calculates positive populations, because the theory has no zero-point energy ($\gamma$) parameter. Although the difference is not as great as for the 300\,K case, nonetheless the absolute errors in the populations are larger than those for spin-PLDM, which again shows itself to be the most accurate of the three methods.

\section{Conclusions}
In this paper we have derived a new partially linearized mapping approach based on classical trajectories, which uses elements of previously derived partially linearized techniques (such as standard PLDM and FBTS) and applies them to the spin-mapping space using the Stratonovich--Weyl transform. We have then tested this method, called spin-PLDM, on a range of model systems in various different regimes. Comparison has also been made with other mapping-based classical-trajectory techniques. 

We show that this method exhibits improved accuracy over standard PLDM, when calculating correlation functions. In particular, spin-PLDM corrects for the large error observed in the long-time limit of identity-containing correlation functions calculated using standard PLDM and also appears to solve the issue of overdamped oscillations observed in many dynamical quantities. This suggests that spin-PLDM is in some sense closer to an exact solution of the QCLE than the standard PLDM method. Spin-PLDM also appears to be able to improve upon previously derived fully linearized approaches, such as spin-LSC\@. In particular, we have observed that spin-PLDM appears to solve the problem of `overdamped coherences' observed in the correlation functions of traceless operators calculated using spin-LSC\@.

For the FMO model, $10^{7}$ trajectories were used to fully converge the focused spin-PLDM results compared to $10^{6}$ trajectories for the other methods; however we also observed that only $10^{4}$ trajectories were needed to qualitatively reproduce the main features of the dynamics. Although we found that focused spin-PLDM generally requires one order of magnitude more trajectories than other mapping-based classical-trajectory methods, we expect that more efficient sampling schemes could easily be implemented in the future in order to reduce the number of required trajectories.\footnote{In particular, 
our current approach was based on uniform sampling of the $F^{2}$ terms within the definition of the spin-PLDM focused sampling. As these terms do not all have equal magnitudes, the number of required trajectories could probably be significantly reduced by using improved sampling schemes which favour the terms with the largest values. Additionally, the prescence of a zero-point energy term means that uniform sampling is clearly not optimal.} In general, our results show that generalized spin-mapping methods outperform their MMST analogues and partially linearized methods appear superior to fully-linearized ones; hence spin-PLDM is the best mapping method of them all. The superior accuracy of spin-PLDM is evident not just for calculating electronic population dynamics, but also for coherences.

In particular, spin-PLDM seems to reproduce the relatively short time properties of real-time quantum correlation functions extremely accurately compared to other mapping-based classical-trajectory techniques. Because of this there are a number of applications which may therefore be well suited for spin-PLDM\@. For instance, memory kernels within the generalized quantum master equation formalism normally decay relatively rapidly and hence it is expected that such quantities can be accurately computed using spin-PLDM, as has been done with other methods.\cite{Shi2004GQME,Montoya2016GQME,*Montoya2017GQME,Mulvihill2019LSCGQME,Kelly2015nonadiabatic,*Kelly2016master,Pfalzgraff2019GQME} Dipole-dipole correlation functions and other optical response functions, from which linear and non-linear spectra can be obtained, also often have short coherence times and hence perhaps can also be accurately computed using spin-PLDM. 

In Paper II,\cite{paper2} the spin-PLDM method is extensively analysed, in order to understand the key differences between spin-PLDM and standard PLDM and between spin-PLDM and spin-LSC. We also outline the implementation of focused initial conditions for spin-PLDM, which were used to generate the results of this work, and introduce a jump spin-PLDM scheme, which in principle allows results to be systematically improved to those of QCLE.
\begin{acknowledgments}
The authors would like to acknowledge the support from the Swiss National Science Foundation through the NCCR MUST (Molecular Ultrafast Science and Technology) Network. We also thank Johan Runeson for helpful discussions and for his comments made on the original manuscript.
\end{acknowledgments}

\section*{Data Availability}
The data that supports the findings of this study are available within the article.

\bibliography{paper1}

\end{document}